\documentclass[preprintnumbers,10pt,nofootinbib]{revtex4}

\usepackage{amsmath,latexsym,amssymb,amsfonts}
\usepackage[pdftex]{color,graphicx}
\usepackage{bm}
\usepackage[normalem]{ulem}

\begin{document}


\title{\textbf{\textit{Void inside the horizon}: connecting the\\ third-quantized vacuum to a classical spacetime}}

\author{
Dong-han Yeom$^{a,b,c}$\footnote{{\tt innocent.yeom@gmail.com}}
}

\affiliation{
$^a$Department of Physics Education, Pusan National University, Busan 46241, Republic of Korea\\
$^b$Research Center for Dielectric and Advanced Matter Physics, Pusan National University, Busan 46241, Republic of Korea\\
$^c$Leung Center for Cosmology and Particle Astrophysics, National Taiwan University, Taipei 10617, Taiwan
}

\begin{abstract}
We investigate the emergence of a classical spacetime from a quantum-gravitational vacuum within the framework of third quantization applied to the black hole interior. Following the canonical quantization of the Kantowski--Sachs spacetime, one can construct a wave packet that explains the classical interior of a Schwarzschild black hole. This wave packet should be extended to the classical singularity. To avoid this, one can introduce an analytic continuation and the Euclidean wave function. This is not an analytic continuation of the spacetime, but of the superspace. To give this a suitable meaning, it is reasonable to introduce the concept of the third quantization. In this aspect, the Euclidean analytic continuation of the superspace explains the on-shell trajectory connecting the third-quantized vacuum state to a coherent classical geometry. Finally, we discuss how nonlinear self-interactions and baby-universe wormhole couplings must be incorporated into the third-quantized action to dynamically drive this transition.
\end{abstract}

\maketitle

\newpage

\tableofcontents

\newpage

\section{Introduction}

Understanding how a classical spacetime manifold emerges from a fundamentally quantum-gravitational regime remains a central problem in theoretical physics. In canonical quantum gravity, the dynamics of geometry and matter are governed by the Wheeler--DeWitt (WDW) equation \cite{DeWitt:1967yk}, which defines a wave function $\Psi[h_{ij}, \phi]$ over superspace. While the WDW framework successfully describes a single quantum geometry, it faces several conceptual challenges.

To avoid the problems of the WDW equation, it has been suggested the idea of the \textit{third quantization} \cite{Giddings:1988wv,McGuigan:1988vi} in which the WDW wave function $\Psi$ is promoted to a quantum field operator $\hat{\Psi}$ acting on a Fock space of universes. Just as second quantization elevates the single-particle wave function to a field operator that describes particle creation and annihilation, third quantization describes the creation of universes from a third-quantized vacuum $|\varnothing\rangle$, so-called the \textit{void}, and their subsequent evolution into macroscopic classical spacetimes.

In this paper, we focus on quantizing the interior of a Schwarzschild black hole \cite{Bouhmadi-Lopez:2019kkt}. The interior of the Schwarzschild black hole is well described by the metric form of anisotropic cosmology \cite{Cavaglia:1994yc}, known as the Kantowski--Sachs spacetime \cite{Kantowski:1966te}. Using this metric, we solve the WDW equation and construct a wave packet that closely follows the classical trajectory \cite{Bouhmadi-Lopez:2019kkt}. However, a problem remains: the wave packet will propagate to the singularity, and it cannot be avoided.

To avoid the central singularity, one may analytically continue to Euclidean time \cite{Chen:2016ask}, following the philosophy of Hartle and Hawking \cite{Hartle:2008ng}. However, inside the horizon, the analytic continuation of the spacetime is not well-defined. The second option is to analytically continue not \textit{space}, but the \textit{superspace} of the WDW wave function~\cite{Yeom:2021bpd}. Interestingly, this analytic continuation can yield a vanishing WDW wave function in the Euclidean domain.

Now one can ask what this means physically. In terms of the WDW wave function, the physical meaning of Euclidean analytic continuation is not very clear in relation to the classical Euclidean geometry. However, if we promote the WDW wave function to the third-quantization scheme, the meaning becomes clearer: the vanishing Euclidean wave function reflects the Euclidean vacuum in the third-quantized field theory, while the analytically continued WDW wave packet emerges as an in--out transition amplitude interpolating between the third-quantized vacuum and a macroscopic coherent state. Thus, in terms of the third-quantized wave function, the wave packet can be interpreted as an analogous \textit{on-shell solution of the superspace}. This new observation will provide profound insights into understanding the quantum gravitational structures of black holes.

This paper is organized as follows. In Sec.~\ref{sec:can}, we discuss the canonical quantization of the Schwarzschild black hole interior. This covers the Wheeler-DeWitt equation and its solutions, as well as the wave packet that explains the classical trajectory. In Sec.~\ref{sec:sev}, we discuss possible scenarios to avoid the singularity. As a result, we propose that Wick rotation to the Euclidean signature might be an interesting way to avoid a singularity. In Sec.~\ref{sec:thi}, we introduce the third quantization scheme, which makes it clear that the Wick rotation is in superspace, not spacetime. Finally, in Sec.~\ref{sec:con}, we summarize our work and discuss possible speculations for future research.

\section{\label{sec:can}Canonical quantization of a Schwarzschild black hole interior}

\subsection{Wheeler--DeWitt equation}

To introduce the canonical formalism for the black hole interior, we adopt the Kantowski--Sachs metric ansatz~\cite{Kantowski:1966te} inside the event horizon:
\begin{equation}
ds^2 = -N^2(t) dt^2 + a^2(t) dr^2 + \frac{r_s^2 b^2(t)}{a^2(t)} d\Omega_2^2,
\label{eq:ks_metric}
\end{equation}
where $N(t)$ is the lapse function, $r_s = 2M$ is the Schwarzschild radius (with the mass $M$), $d\Omega_2^2 = d\theta^2 + \sin^2\theta\, d\varphi^2$ is the unit two-sphere metric, and $a(t), b(t)$ are the two canonical metric variables. 

The Schwarzschild solution inside the event horizon is presented as
\begin{equation}
ds^{2} = - \left( \frac{2M}{t} - 1 \right)^{-1} dt^{2} + \left( \frac{2M}{t} - 1 \right) dr^{2} + t^{2} d\Omega_{2}^{2}.
\end{equation}
By comparing this solution with Eq.~\eqref{eq:ks_metric} and eliminating the $t$ parameter, regardless of the gauge choice for $N(t)$, the classical Schwarzschild interior solution satisfies the on-shell algebraic relation:
\begin{equation}
\frac{1}{b} = a + \frac{1}{a}.
\label{eq:classical_onshell}
\end{equation}

\begin{figure}[t]
\centering
\includegraphics[width=0.95\textwidth]{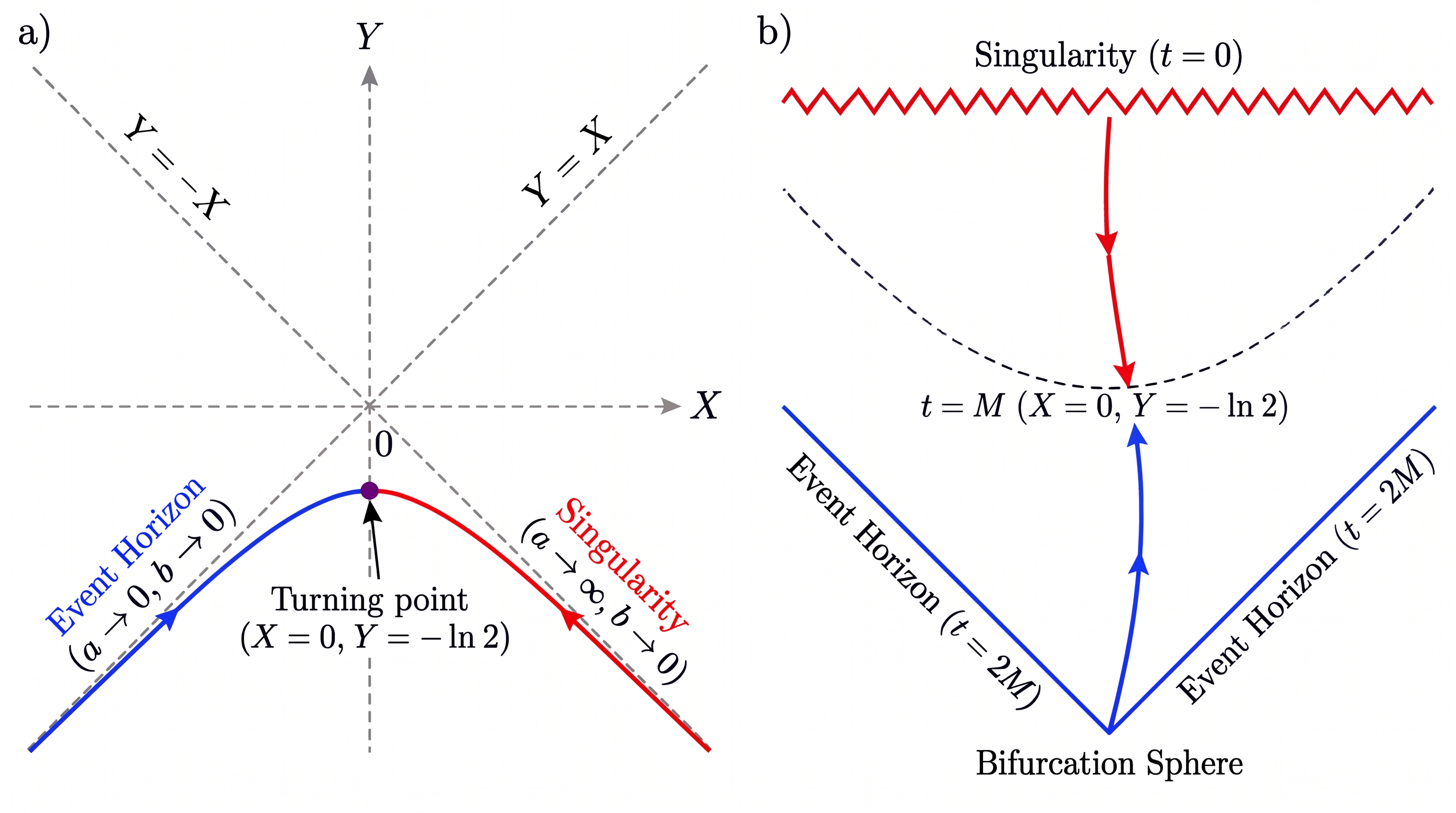}
\caption{Classical on-shell trajectory in minisuperspace and its spacetime correspondence inside the Schwarzschild black hole horizon. 
(a) The classical solution $Y(X) = -\ln(2\cosh X)$ in the $(X, Y)$ minisuperspace plane. The blue asymptotic branch ($X \simeq Y \to -\infty$) corresponds to the event horizon ($a \to 0, b \to 0$), the turning point occurs at $(X = 0, Y = -\ln 2)$, and the red asymptotic branch ($X \simeq -Y \to +\infty$) approaches the classical singularity ($a \to \infty, b \to 0$). 
(b) The causal Penrose diagram of the Schwarzschild black hole interior, illustrating the corresponding evolution between the event horizon ($t = 2M$, blue), the turning-point hypersurface ($t = M$, dashed), and the spacelike singularity ($t = 0$, red).}
\label{fig:classical_minisuperspace}
\end{figure}

Defining logarithmic minisuperspace coordinates $X \equiv \ln a$ and $Y \equiv \ln b$, the classical on-shell trajectory, Eq.~\eqref{eq:classical_onshell}, translates into
\begin{equation}
e^{-Y} = e^{X} + e^{-X} \quad \Longleftrightarrow \quad Y(X) = -\ln\left(e^{X} + e^{-X}\right) = -\ln\left(2\cosh X\right).
\label{eq:onshell_XY}
\end{equation}
As illustrated in Fig.~\ref{fig:classical_minisuperspace}(a), in the $(X, Y)$ plane, the event horizon ($a \to 0, b \to 0$) corresponds to the asymptotic limit $X \simeq Y \to -\infty$, the classical turning point in minisuperspace occurs at $X = 0$ ($Y = -\ln 2$), and the classical singularity ($a \to \infty, b \to 0$) corresponds to $X \simeq -Y \to +\infty$. Fig.~\ref{fig:classical_minisuperspace}(b) shows the corresponding causal structure inside the event horizon ($t = 2M$), where the interior geometry evolves across the $t = M$ hypersurface ($X = 0$) toward the spacelike singularity ($t = 0$).

Substituting the metric ansatz, Eq.~\eqref{eq:ks_metric}, into the vacuum Einstein--Hilbert action and performing canonical quantization yields the minisuperspace Wheeler--DeWitt equation \cite{Bouhmadi-Lopez:2019kkt}:
\begin{equation}
\left( \frac{\partial^2}{\partial X^2} - \frac{\partial^2}{\partial Y^2} + 4r_s^2 e^{2Y} \right) \Psi(X, Y) = 0,
\label{eq:wdw_bh}
\end{equation}
where $X$ plays the role of a time-like coordinate and $Y$ acts as a space-like coordinate subject to an exponential potential barrier $V(Y) = 4r_s^2 e^{2Y}$ that diverges as $Y \to +\infty$.

\subsection{Wave packets}

Using separation of variables $\Psi(X, Y) = \varphi(X)\psi(Y)$ with separation constant $k^2 > 0$, Eq.~\eqref{eq:wdw_bh} decouples into
\begin{equation}
\left( \frac{d^2}{dX^2} + k^2 \right) \varphi(X) = 0, \qquad \left( \frac{d^2}{dY^2} - 4r_s^2 e^{2Y} + k^2 \right) \psi(Y) = 0.
\end{equation}
Imposing boundedness as $Y \to +\infty$ eliminates the growing modified Bessel function $I_{ik}(2r_s e^Y)$, leaving only the modified Bessel function of the second kind $K_{ik}(2r_s e^Y)$. Thus, the general solution takes the integral form:
\begin{equation}
\Psi(X, Y) = \int_{-\infty}^{\infty} f(k) e^{-ikX} K_{ik}\left(2r_s e^Y\right) dk.
\label{eq:general_bessel}
\end{equation}

\begin{figure}[t]
\centering
\includegraphics[width=0.78\textwidth]{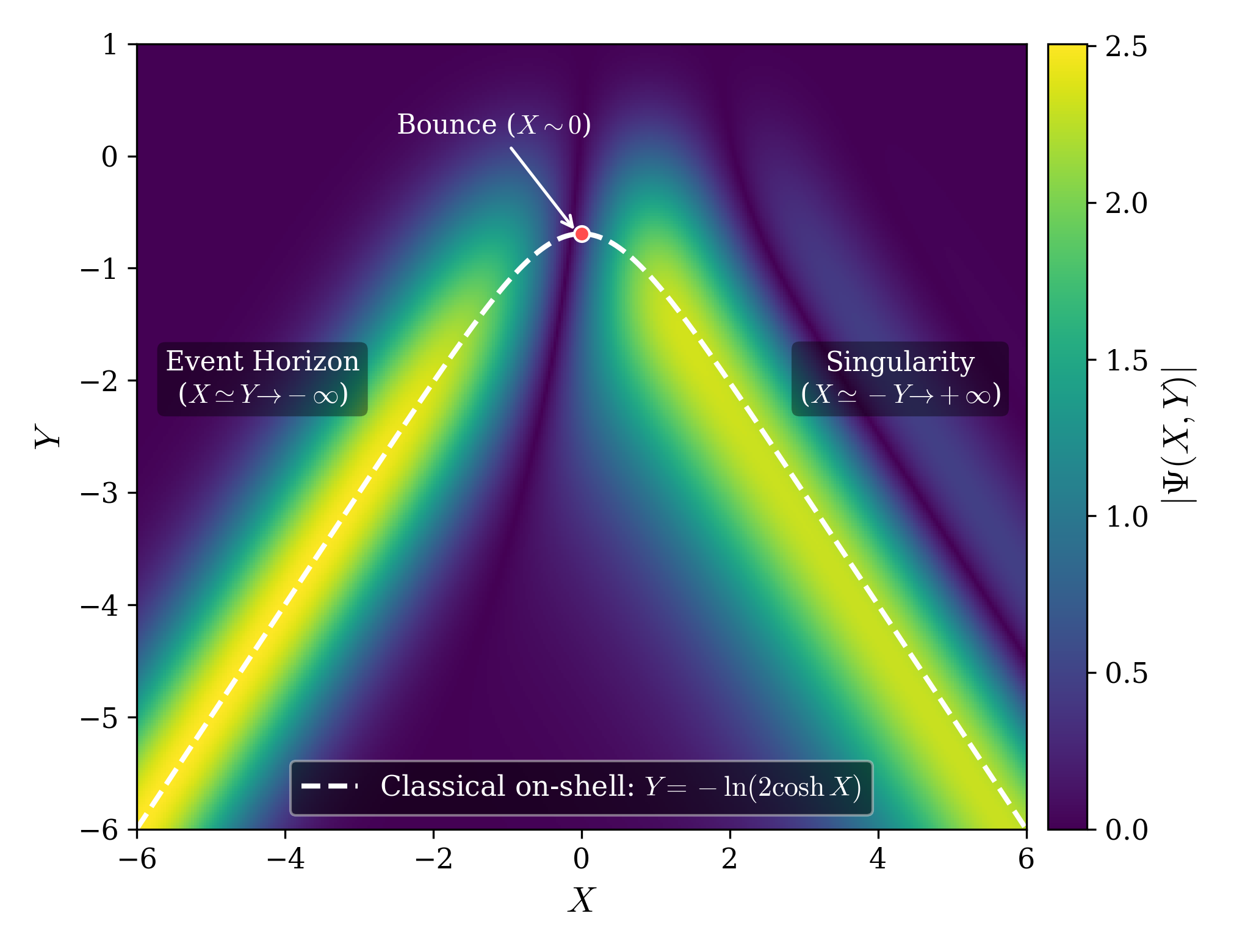}
\caption{Numerical evaluation of the Lorentzian Wheeler--DeWitt wave packet amplitude $|\Psi(X, Y)|$ given by Eqs.~\eqref{eq:general_bessel} and~\eqref{eq:gaussian_fk} for $A = 1$, $\sigma = 1$, and $r_s = 1$. The white dashed curve represents the classical Schwarzschild interior trajectory $Y(X) = -\ln(2\cosh X)$, with the red dot marking the classical turning (bounce) point at $(X = 0, Y = -\ln 2)$. The incoming packet closely tracks the classical path near the event horizon ($X \simeq Y \to -\infty$), undergoes a quantum bounce against the exponential potential barrier $4r_s^2 e^{2Y}$ around $X \sim 0$, and propagates along the reflected branch toward the singularity ($X \simeq -Y \to +\infty$) with mild wave-packet dispersion induced by the scattering phase $\Gamma(ik)/\Gamma(-ik)$.}
\label{fig:wdw_wavepacket_2d}
\end{figure}

Near the event horizon ($Y \to -\infty$), the Bessel function admits the asymptotic expansion:
\begin{equation}
K_{ik}\left(2r_s e^Y\right) \simeq \frac{1}{2} \left( r_s^{ik} e^{ikY} \Gamma(-ik) + r_s^{-ik} e^{-ikY} \Gamma(ik) \right).
\label{eq:bessel_asymp}
\end{equation}
To satisfy Ehrenfest's theorem so that the wave function peaks along the classical on-shell trajectory $X = Y$ near the horizon ($X, Y \to -\infty$), we select the spectral weight function \cite{Bouhmadi-Lopez:2019kkt}:
\begin{equation}
f(k) = \frac{2A e^{-\sigma^2 k^2 / 2}}{\Gamma(-ik) r_s^{ik}},
\label{eq:gaussian_fk}
\end{equation}
where $A$ is a normalization constant and $\sigma$ is the standard deviation of the pulse. Substituting Eqs.~\eqref{eq:bessel_asymp} and~\eqref{eq:gaussian_fk} into Eq.~\eqref{eq:general_bessel} in the horizon limit ($Y \to -\infty$) yields two interfering Gaussian wave packets:
\begin{align}
\Psi(X, Y) &\simeq A \int_{-\infty}^{\infty} e^{-\sigma^2 k^2 / 2} \left[ e^{-ik(X - Y)} + \frac{\Gamma(ik)}{\Gamma(-ik)} r_s^{-2ik} e^{-ik(X + Y)} \right] dk \nonumber \\
&\simeq A \sqrt{\frac{2\pi}{\sigma^2}} \exp\left[ -\frac{(X - Y)^2}{2\sigma^2} \right] + \Psi_{\mathrm{reflected}}(X + Y).
\label{eq:horizon_gaussian}
\end{align}
The first term represents an incoming Gaussian wave packet localized sharply along the classical on-shell path $X = Y$ ($Y \to -\infty$). As this packet propagates toward larger $X$, it encounters the exponential potential wall $4r_s^2 e^{2Y}$ near $X \sim 0$, undergoing a quantum bounce into the second (reflected) branch along $X = -Y$ toward the singularity.

Fig.~\ref{fig:wdw_wavepacket_2d} illustrates the numerical profile of $|\Psi(X, Y)|$ for $A = \sigma = r_s = 1$, clearly showing how the incoming wave packet follows the classical on-shell curve near the horizon before bouncing near $X \sim 0$ and extending toward the singular regime.

\section{\label{sec:sev}Several ways toward singularity avoidance}

\subsection{Revisit several scenarios}

This wave packet covers the singularity ($X \rightarrow \infty$ and $Y \rightarrow -\infty$). To avoid the singularity, there might be several ways:
\begin{itemize}
\item[--] 1. If we impose a naive vanishing boundary condition near the singularity, we need to introduce the modified Bessel function $I_{ik}(2r_s e^Y)$, which leads to unavoidable divergence \cite{Bouhmadi-Lopez:2019kkt}.
\item[--] 2. If we interpret the two arrows of time, one can introduce destructive interference at $r \sim M$, called \textit{annihilation-to-nothing} \cite{Bouhmadi-Lopez:2019kkt,Brahma:2021xjy,Chien:2023kqw}. This explains the annihilation of two spacetime patches, where one patch should include a (white-hole-like) singularity anyway.
\item[--] 3. Introducing a gauge field may yield a wave packet that vanishes nicely in the interior, even for an averagely neutral black hole. However, this relies on the existence of a gauge field \cite{Chien:2025tzm}.
\end{itemize}
Therefore, it is reasonable to seek a mechanism that imposes a vanishing boundary condition inside the black hole and can be applied to generic neutral black holes to avoid the singularity.

\subsection{Introducing Wick rotations}

\subsubsection{Wick rotation of space}

To introduce the Wick rotation, the first reasonable attempt is to choose an arbitrary time slice \cite{Chen:2016ask}: for a given Schwarzschild solution ($r < 2M$)
\begin{equation}
ds^{2} = \left( \frac{2M}{r} - 1 \right)dt^{2} - \left(\frac{2M}{r} - 1\right)^{-1} dr^{2} + r^{2} d\Omega_{2}^{2},
\end{equation}
we Wick rotate $r = r_{0} - i \rho$. The resulting metric is
\begin{equation}
ds^{2} = \left( \frac{2M}{r_{0} - i\rho} - 1 \right)dt^{2} + \left(\frac{2M}{r_{0} - i\rho} - 1\right)^{-1} d\rho^{2} + (r_{0} - i\rho)^{2} d\Omega_{2}^{2}.
\end{equation}
This metric is apparently complex-valued, but in the large $\rho$ limit, the metric approaches
\begin{equation}
ds^{2} = - \left( dt^{2} + d\rho^{2} + \rho^{2} d\Omega_{2}^{2} \right),
\end{equation}
and hence, it asymptotically approaches Euclidean geometry with an overall minus sign relative to the $(-,+,+,+)$ Lorentzian signature. However, although the bulk Ricci scalar vanishes identically for the vacuum solution, this Euclideanized domain has no horizon to fix the Euclidean periodicity or the Gibbons--Hawking--York boundary contribution unambiguously \cite{Gibbons:1976ue}. Consequently, introducing a well-defined probability interpretation is difficult, and this analytic continuation has limited theoretical utility.

\subsubsection{Wick rotation of superspace}
\label{sec:wick_superspace}

Another theoretical possibility is to introduce the Wick rotation in the superspace. The separation variable $k$ can be interpreted as an energy parameter if we interpret $X$ as a time-like variable. If we take this more seriously as a time-like evolution, we need to select only the positive-$k$ domain; this is harmless, as it matches the correspondence between the wave packet and the classical trajectory.

\begin{figure}[t]
\centering
\includegraphics[width=0.52\textwidth]{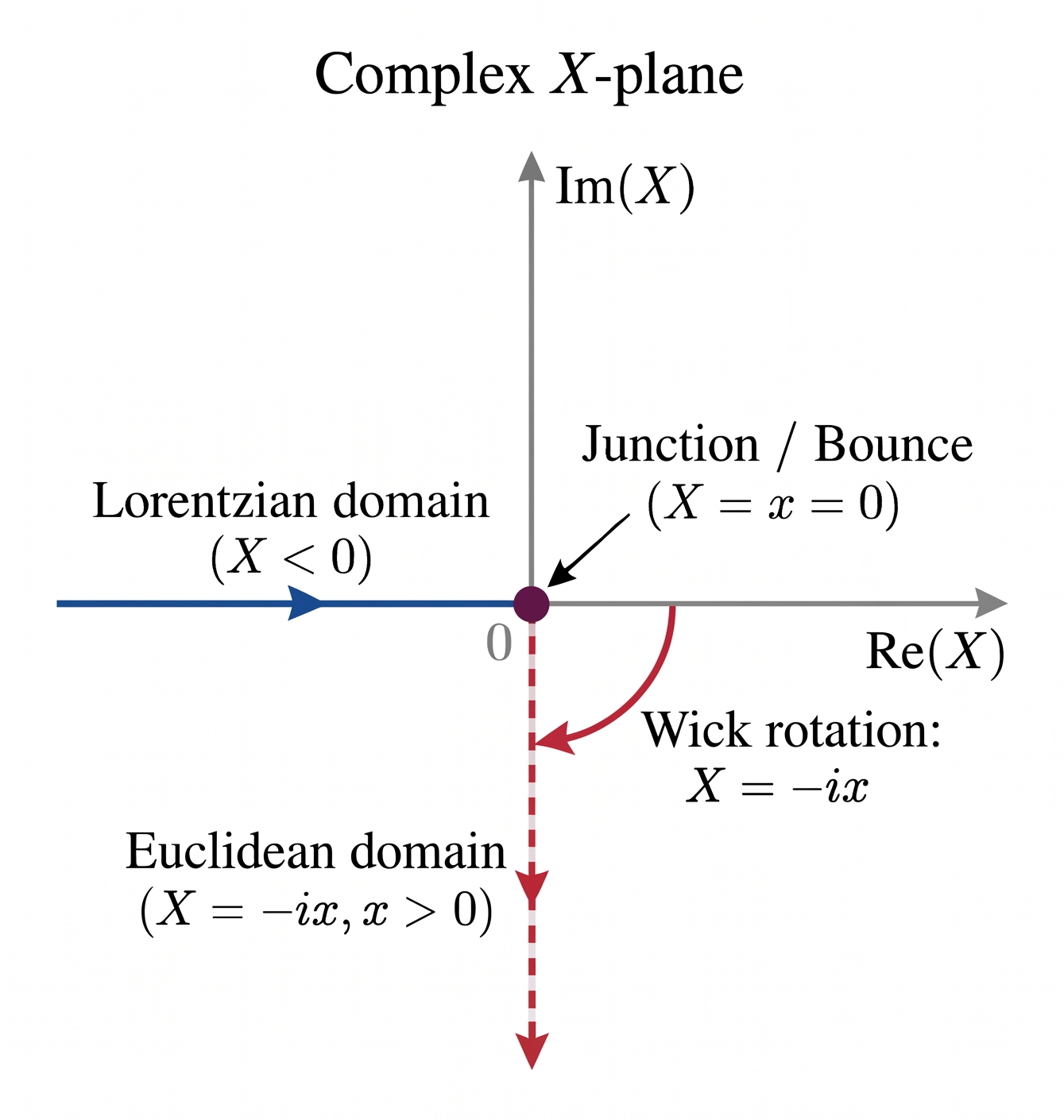}
\caption{Analytic continuation contour in the complex $X$-plane illustrating the minisuperspace Wick rotation at the junction hypersurface $X = x = 0$. The Lorentzian interior evolution proceeds along the negative real axis ($X < 0$, blue solid line) up to the turning point $X = 0$, where a clockwise $90^\circ$ Wick rotation $X = -ix$ ($x > 0$) connects the trajectory onto the negative imaginary axis corresponding to the Euclidean domain (red dashed line).}
\label{fig:wick_contour}
\end{figure}

Since $X$ is the time-like parameter in the superspace, it is not surprising to introduce the following analytic continuation: we now introduce a Euclidean analytical continuation of the time-like minisuperspace coordinate $X$ at the bouncing hypersurface $X = 0$:
\begin{equation}
X = -i x, \qquad x > 0.
\label{eq:wick}
\end{equation}
As illustrated in Fig.~\ref{fig:wick_contour}, performing the clockwise Wick rotation $X = -ix$ ($x > 0$) at the bounce junction $X = x = 0$ analytically continues the Lorentzian domain ($X < 0$) onto the negative imaginary axis of the complex $X$-plane.

Under this Wick rotation, the oscillatory factor $e^{-ikX}$ inside the integral Eq.~\eqref{eq:general_bessel} transforms into $e^{-kx}$. Thus, the Lorentzian wave packet for $X < 0$ is defined as
\begin{equation}
\Psi(X, Y) = \int_{0}^{\infty} f(k) e^{-ikX} K_{ik}\left(2r_s e^Y\right) dk \qquad (X < 0),
\label{eq:lorentz_kpos}
\end{equation}
and upon analytical continuation via $X = -ix$ across $X = 0$, the Euclidean wave function for $x > 0$ becomes
\begin{equation}
\Psi(x, Y) = \int_{0}^{\infty} f(k) e^{-kx} K_{ik}\left(2r_s e^Y\right) dk \qquad (x > 0),
\label{eq:euclid_kpos}
\end{equation}
with $f(k)$ given by Eq.~\eqref{eq:gaussian_fk} \cite{Yeom:2021bpd}.

\begin{figure}[t]
\centering
\includegraphics[width=0.80\textwidth]{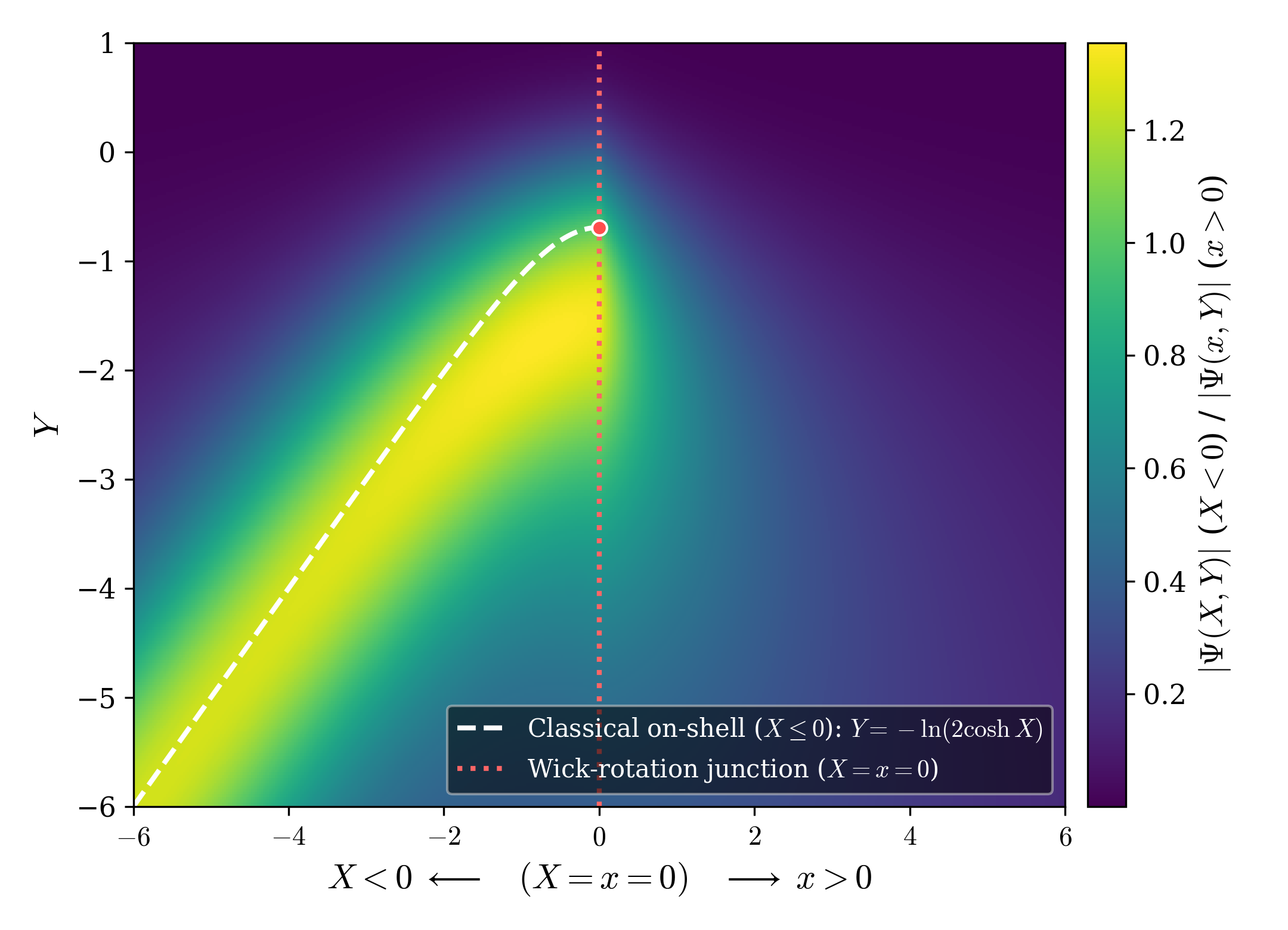}
\caption{Numerical evaluation of the positive-$k$ ($k > 0$) Wheeler--DeWitt wave packet across the Wick-rotation junction $X = x = 0$ (red dotted line) for $A = \sigma = r_s = 1$. In the Lorentzian domain ($X < 0$), $|\Psi(X, Y)|$ follows the classical Schwarzschild interior trajectory $Y(X) = -\ln(2\cosh X)$ (white dashed curve) near the horizon before approaching the turning region around $X \sim 0$. For $x > 0$ under $X = -ix$, $|\Psi(x, Y)|$ decays exponentially as $x \to \infty$ without forming a real Euclidean steepest-descent ridge.}
\label{fig:wick_numerical_clean}
\end{figure}

This wave function satisfies the following properties (Fig.~\ref{fig:wick_numerical_clean}):
\begin{itemize}
    \item[--] \textit{Lorentzian domain}: Even after restricting the integration domain to $k > 0$ in Eq.~\eqref{eq:lorentz_kpos}, the half-Gaussian integral still exhibits a prominent ridge peaking along the classical Schwarzschild on-shell trajectory $Y(X) = -\ln(2\cosh X) \simeq X$ near the event horizon ($X \ll -1$), satisfying Ehrenfest's theorem in the asymptotic Lorentzian regime. Only near the bouncing region $X \sim 0$, quantum interference and the $k > 0$ truncation induce a noticeable deviation from the classical curve.
    
    \item[--] \textit{Euclidean domain}: Across $X = 0$ into the $x > 0$ region governed by Eq.~\eqref{eq:euclid_kpos}, the damping factor $e^{-kx}$ suppresses all $k > 0$ modes, causing the wave packet amplitude $|\Psi(x, Y)|^2$ to decay rapidly to zero\footnote{Indeed, the Wick rotation need not occur at $X = 0$; once the continuation is performed on any surface, the Euclidean suppression factor appears.}:
    \begin{equation}
    \lim_{x \to \infty} \Psi(x, Y) = 0.
    \end{equation}
    Moreover, high-$k$ (high-energy) modes decay fastest, so the $x \to \infty$ limit dynamically projects out excited modes and isolates the lowest-energy state. After the Wick rotation, the corresponding on-shell contour condition satisfies $Y(x) = -\ln(2 \cos x)$. However, the Euclidean region ($x > 0$) of $|\Psi(x, Y)|^2$ shows no localized steepest-descent ridge (i.e., no well-defined Euclidean on-shell trajectory).
\end{itemize}

\section{\label{sec:thi}Third quantization of the black hole interior}

\subsection{Third quantization program}

We now elevate the single-universe WDW wave packet of Eqs.~\eqref{eq:lorentz_kpos} and~\eqref{eq:euclid_kpos} to the framework of third quantization \cite{Giddings:1988wv,McGuigan:1988vi}. In this formalism, the WDW equation, Eq.~\eqref{eq:wdw_bh}, arises as the classical Euler--Lagrange equation of a third-quantized field action on the $(X, Y)$ minisuperspace:
\begin{equation}
\mathcal{S}[\Psi] = \frac{1}{2} \int dX \, dY \left[ \left(\frac{\partial \Psi}{\partial X}\right)^2 - \left(\frac{\partial \Psi}{\partial Y}\right)^2 - 4r_s^2 e^{2Y}\Psi^2 \right].
\label{eq:3q_action}
\end{equation}
Promoting $\Psi(X, Y)$ to a quantum field operator $\hat{\Psi}(X, Y)$ with canonical commutation relations
\begin{equation}
[\hat{\Psi}(X,Y), \hat{\Pi}_\Psi(X,Y')] = i\delta(Y - Y'),
\end{equation}
we expand $\hat{\Psi}$ in terms of the exact normal modes
\begin{equation}
u_k(X, Y) = \frac{1}{\sqrt{2k}} e^{-ikX} \tilde{K}_{ik}\left(2r_s e^Y \right),
\end{equation}
where
\begin{equation}
\tilde{K}_{ik}\left(2r_s e^Y\right) \equiv \frac{\sqrt{2k\sinh(\pi k)}}{\pi} K_{ik}\left(2r_s e^Y\right)
\end{equation}
is the real, delta-function-normalized Kontorovich--Lebedev Bessel eigenfunction satisfying
\begin{equation}
\int_{-\infty}^{\infty} dY \, \tilde{K}_{ik}(2r_s e^Y) \tilde{K}_{ik'}(2r_s e^Y) = \delta(k - k')
\end{equation}
for $k, k' > 0$:
\begin{equation}
\hat{\Psi}(X, Y) = \int_{0}^{\infty} dk \left( \hat{A}_k u_k(X, Y) + \hat{A}_k^\dagger u_k^*(X, Y) \right),
\label{eq:3q_field}
\end{equation}
where $[\hat{A}_k, \hat{A}_{k'}^\dagger] = \delta(k - k')$. Here, $\hat{A}_k^\dagger$ creates a single quantum black-hole-interior universe in mode $k$ from the third-quantized vacuum $|\varnothing\rangle$.

\subsection{Third-quantized Euclidean vacuum: \textit{Euclidean void}}

Performing the Wick rotation $X = -ix$ ($x > 0$) on the mode expansion, Eq.~\eqref{eq:3q_field}, transforms the temporal factors as $e^{-ikX} \to e^{-kx}$ and $e^{+ikX} \to e^{+kx}$:
\begin{equation}
\hat{\Psi}(x, Y) = \int_{0}^{\infty} \frac{dk}{\sqrt{2k}} \left( \hat{A}_k e^{-kx} + \hat{A}_k^\dagger e^{+kx} \right) \tilde{K}_{ik}\left(2r_s e^Y \right).
\label{eq:3q_euclid_field}
\end{equation}

Notice that for $k>0$, as $x\to+\infty$, the term multiplying $\hat{A}_k$
decays exponentially ($e^{-kx}\to 0$), whereas the term multiplying
$\hat{A}^\dagger_k$ grows exponentially ($e^{+kx}\to\infty$).
Since the one-point function $\langle s |\hat{\Psi}| s \rangle$ vanishes
identically for any Fock state of the free theory, the vacuum must be
selected at the level of correlation functions. We therefore require that
the Euclidean two-point function remain bounded in the asymptotic Euclidean
limit:
\begin{equation}
\lim_{x\to+\infty}
\big\langle s \big|\hat{\Psi}(x,Y)\,\hat{\Psi}(x',Y')\big| s \big\rangle
< \infty
\qquad (x > x' > 0).
\label{eq:regularity}
\end{equation}
For a generic state $|s \rangle$, the contribution
$\langle s |\hat{A}^\dagger_k \hat{A}_{k'}| s \rangle\,
e^{+kx}e^{-k'x'}$ grows without bound, so
Eq.~(\ref{eq:regularity}) requires
$\langle s |\hat{A}^\dagger_k \hat{A}_k| s \rangle
= \big\|\hat{A}_k| s \rangle\big\|^2 = 0$ for all $k>0$.
This uniquely selects the Euclidean vacuum $|\varnothing_E\rangle$:
\begin{equation}
\hat{A}_k|\varnothing_E\rangle = 0 \qquad (\forall\, k>0),
\label{eq:vacuum}
\end{equation}
for which the two-point function becomes
\begin{equation}
\big\langle \varnothing_E\big|\hat{\Psi}(x,Y)\,\hat{\Psi}(x',Y')
\big|\varnothing_E\big\rangle
= \int_0^\infty \frac{dk}{2k}\, e^{-k(x-x')}\,
\tilde{K}_{ik}\!\left(2r_s e^{Y}\right)
\tilde{K}_{ik}\!\left(2r_s e^{Y'}\right),
\label{eq:twopoint}
\end{equation}
which decays exponentially as $x\to+\infty$.
Equivalently, the superspace Wick rotation acts as an $i\epsilon$
prescription: Euclidean evolution along $x$ projects any state onto the
positive-frequency Fock vacuum defined with respect to the time-like
minisuperspace variable $X$. This also explains why the location of the
Wick-rotation surface is irrelevant (see footnote~1).
Therefore, the Euclidean boundary condition $\Psi(x \to \infty, Y) \to 0$ imposed in Sec.~\ref{sec:wick_superspace} is precisely the regularity condition defining the third-quantized vacuum $|\varnothing_E\rangle$—the state of absolute ``nothingness'' prior to the nucleation of a classical interior geometry, so to speak, the \textit{Euclidean void} state.

\subsection{Coherent states and the on-shell transition amplitude}

To describe a macroscopic classical black hole interior, we act on $|\varnothing_E\rangle$ with the unitary displacement operator $\hat{D}[\tilde{f}]$ to form a third-quantized coherent state $|\Psi_{\mathrm{cl}}\rangle$:
\begin{equation}
|\Psi_{\mathrm{cl}}\rangle = \hat{D}[\tilde{f}] |\varnothing_E\rangle = \exp\left( \int_{0}^{\infty} dk \left[ \tilde{f}(k)\hat{A}_k^\dagger - \tilde{f}^*(k)\hat{A}_k \right] \right) |\varnothing_E\rangle,
\end{equation}
where $\tilde{f}(k)$ is chosen to absorb both the $1/\sqrt{2k}$ mode normalization and the $\tilde{K}_{ik}$ factor so that $\tilde{f}(k) u_k(X, Y) = f(k) e^{-ikX} K_{ik}(2r_s e^Y)$, namely
\begin{equation}
\tilde{f}(k) = \frac{\pi}{\sqrt{\sinh(\pi k)}} f(k) = \frac{2\pi A e^{-\sigma^2 k^2 / 2}}{\sqrt{\sinh(\pi k)}\, \Gamma(-ik) r_s^{ik}}.
\label{eq:tilde_fk}
\end{equation}
Using the Gamma-function identity $|\Gamma(-ik)|^2 = \pi / [k \sinh(\pi k)]$, we find
\begin{equation}
|\tilde{f}(k)|^2 = 4\pi A^2 k e^{-\sigma^2 k^2},
\end{equation}
which is strictly square-integrable:
\begin{equation}
\int_0^\infty |\tilde{f}(k)|^2 dk = 2\pi A^2 / \sigma^2 < \infty.
\end{equation}
Thus $|\Psi_{\mathrm{cl}}\rangle$ is a well-defined coherent state in the Fock space built upon $|\varnothing_E\rangle$, with a nonzero vacuum overlap $\langle \varnothing_E | \Psi_{\mathrm{cl}} \rangle = \exp(-\pi A^2 / \sigma^2)$.

Because the diagonal expectation value $\langle \Psi_{\mathrm{cl}} | \hat{\Psi}(x, Y) | \Psi_{\mathrm{cl}} \rangle$ involves $\langle \Psi_{\mathrm{cl}} | \hat{A}_k^\dagger | \Psi_{\mathrm{cl}} \rangle = \tilde{f}^*(k)$ multiplying the growing factor $e^{+kx}$ in Eq.~\eqref{eq:3q_euclid_field} (which would restore $k < 0$ modes and diverge as $x \to +\infty$), the analytically continued single-universe WDW wave packet is instead identified with the normalized \textit{in--out} transition matrix element between $|\Psi_{\mathrm{cl}}\rangle$ and $\langle \varnothing_E |$. Since $\hat{A}_k |\Psi_{\mathrm{cl}}\rangle = \tilde{f}(k)|\Psi_{\mathrm{cl}}\rangle$ and $\langle \varnothing_E | \hat{A}_k^\dagger = 0$, this transition amplitude evaluates directly to the exact positive-$k$ wave packet Eq.~\eqref{eq:lorentz_kpos} in the Lorentzian domain ($X < 0$):
\begin{equation}
\Psi_{\mathrm{cl}}(X, Y) \equiv \frac{\langle \varnothing_E | \hat{\Psi}(X, Y) | \Psi_{\mathrm{cl}} \rangle}{\langle \varnothing_E | \Psi_{\mathrm{cl}} \rangle} = \int_{0}^{\infty} dk \, f(k) e^{-ikX} K_{ik}\left(2r_s e^Y\right),
\label{eq:coherent_vev}
\end{equation}
and under $X = -ix$ ($x > 0$), it reduces identically to the decaying Euclidean wave function Eq.~\eqref{eq:euclid_kpos}:
\begin{equation}
\Psi_{\mathrm{cl}}(x, Y) \equiv \frac{\langle \varnothing_E | \hat{\Psi}(x, Y) | \Psi_{\mathrm{cl}} \rangle}{\langle \varnothing_E | \Psi_{\mathrm{cl}} \rangle} = \int_{0}^{\infty} dk \, f(k) e^{-kx} K_{ik}\left(2r_s e^Y\right).
\label{eq:coherent_euclid_vev}
\end{equation}
Because the free third-quantized action, Eq.~\eqref{eq:3q_action}, is quadratic, the field equation holds as an operator identity, and the states $\langle \varnothing_E |$ and $|\Psi_{\mathrm{cl}}\rangle$ are independent of $(X,Y)$ in the Heisenberg picture. Hence, Ehrenfest's theorem holds identically for the in--out matrix element as well: $\Psi_{\mathrm{cl}}(X, Y)$ is an exact (complex) saddle-point solution of the WDW equation, Eq.~\eqref{eq:wdw_bh}, whose amplitude peak tracks the classical Schwarzschild interior trajectory near the horizon.

\subsection{The wave packet as an on-shell solution of the third-quantized wave function}

This correspondence provides a natural, unified interpretation of the wave function. Even without specifying the microscopic dynamical mechanism that excites the field out of the vacuum, the wave packet~$\Psi_{\mathrm{cl}}$—interpolating from the Euclidean branch, Eq.~\eqref{eq:euclid_kpos} ($x > 0$), to the Lorentzian branch, Eq.~\eqref{eq:lorentz_kpos} ($X < 0$)—represents a \textit{classical complex field trajectory} on minisuperspace from the perspective of third quantization:
\begin{itemize}
    \item[--] \textit{Asymptotic Euclidean boundary} ($x \to \infty$): The transition amplitude decays to zero ($\Psi_{\mathrm{cl}} \to 0$), matching the boundary condition of the third-quantized vacuum $|\varnothing_E\rangle$ (``nothing''). As noted in Sec.~\ref{sec:wick_superspace}, no real classical Euclidean on-shell trajectory exists in this regime; rather, the decay of $\Psi_{\mathrm{cl}}(x, Y)$ represents a quantum-gravitational tunneling tail in superspace.
    
    \item[--] \textit{Nucleation and bounce surface} ($x = X = 0$): Across the Wick-rotation junction, the field undergoes a semiclassical transition, where the quantum bounce near the potential wall $4r_s^2 e^{2Y}$ connects smoothly to the decaying Euclidean tail.
    
    \item[--] \textit{Asymptotic Lorentzian regime} ($X \to -\infty$): The transition amplitude settles into a localized half-Gaussian wave packet ($\Psi_{\mathrm{cl}} \neq 0$) whose amplitude peak tracks the classical Schwarzschild on-shell trajectory $b^{-1} = a + a^{-1}$ ($Y = -\ln(2\cosh X)$).
\end{itemize}
Thus, in the third-quantized picture, the analytically continued WDW wave packet serves as the semiclassical bridge connecting the third-quantized vacuum of the ``void'' to a classical black hole interior spacetime. In causal terms, the wave packet on the hypersurface near the event horizon evolves into the Euclidean void in the third-quantized point of view.

\subsection{Roles of nonlinear interactions}

Our explicit WDW wave packet ensures that an on-shell trajectory connects the event horizon to the Euclidean void. This strongly suggests that a unitary description of the third-quantized wave function must exist to explain this on-shell trajectory.

The question is, what forces justify the excited state of the void? To dynamically realize the transition between $|\varnothing_E\rangle$ and $|\Psi_{\mathrm{cl}}\rangle$ without imposing the coherent state \textit{ad hoc}, one must introduce \textit{nonlinear interaction terms} into the third-quantized action:
\begin{equation}
\mathcal{S}_{\mathrm{int}}[\Psi] = -\int dX \, dY \left( \frac{\lambda_3}{3!} \Psi^3 + \frac{\lambda_4}{4!} \Psi^4 + \cdots \right).
\end{equation}
Physically, these nonlinear couplings correspond to topology-changing processes in quantum gravity, such as the branching of baby universes or Euclidean wormholes \cite{Coleman:1988cy}. When microscopic baby-universe modes are integrated out or projected onto wormhole $\alpha$-eigenstates \cite{Coleman:1988tj}, the cubic interaction $\lambda_3 \hat{\Psi}^3$ induces an effective linear source term $J(X,Y)\hat{\Psi}(X,Y)$ in the macroscopic Hamiltonian.

Provided the Euclidean continuation of this effective source $J(x,Y)$ is localized near the bounce surface ($x \sim 0$) and vanishes asymptotically ($x \to +\infty$), it drives the vacuum $|\varnothing_E\rangle$ directly into a coherent state $|\Psi_{\mathrm{cl}}\rangle$ via a finite-action saddle point. Finally, depending on the physical orientation of time inside the horizon, this same superspace instanton admits dual causal interpretations: from the cosmological/nucleation viewpoint, a classical interior emerges out of the void; conversely, aligning the superspace transition with the infalling arrow of time—where a classical geometry evolving from $X \to -\infty$ encounters the quantum bounce near $X \sim 0$ and tunnels into the asymptotic Euclidean void ($x \to +\infty$)—provides a consistent realization of the black hole final-state projection \cite{Horowitz:2003he} and singularity resolution \cite{Bouhmadi-Lopez:2019kkt,Brahma:2021xjy,Chien:2023kqw,Chien:2025tzm}. In any case, constructing the source term from more microscopic principles is beyond the scope of this paper, and we postpone it to future research.

\section{\label{sec:con}Conclusion}

\subsection{Summary}

We have demonstrated that the third-quantized framework resolves both the boundary-condition ambiguity and the conceptual interpretation of superspace analytic continuation inside the Schwarzschild black hole horizon. Rather than performing an ill-defined Wick rotation of the interior spacetime coordinates, analytically continuing the time-like minisuperspace variable $X = -ix$ ($x > 0$) transforms the third-quantized mode expansion into exponentially growing and decaying Euclidean sectors. Requiring regularity as $x \to \infty$ uniquely singles out the third-quantized Euclidean vacuum $|\varnothing_E\rangle$, thereby elevating the ad hoc vanishing boundary condition of the single-universe WDW wave function to a fundamental vacuum-selection principle.

By evaluating the normalized in--out transition matrix element $\langle \varnothing_E | \hat{\Psi} | \Psi_{\mathrm{cl}} \rangle / \langle \varnothing_E | \Psi_{\mathrm{cl}} \rangle$ for a coherent state $|\Psi_{\mathrm{cl}}\rangle$ constructed on $|\varnothing_E\rangle$, we recovered the exact positive-$k$ Lorentzian wave packet tracking the classical Kantowski--Sachs trajectory near the horizon, together with its smooth decay into the Euclidean void beyond the bounce hypersurface $X = 0$. Consequently, the classical black hole interior and the quantum-gravitational ``nothingness'' replacing the central singularity are unified as two asymptotic regimes of a single semiclassical trajectory in minisuperspace, dynamically mediated by wormhole-induced nonlinearities in the third-quantized action.

\subsection{Speculations}

Several intriguing directions emerge from this third-quantized picture of the black hole interior:
\begin{itemize}
    \item[--] \textit{Information paradox and final-state projection}: Interpreting the transition from $|\Psi_{\mathrm{cl}}\rangle$ ($X < 0$) to $|\varnothing_E\rangle$ ($x \to +\infty$) as a dynamical final-state boundary condition provides a natural field-theoretic arena for studying black hole unitarity. This description suggests that the wave function is universally connected to the Euclidean void in third quantization. Does this realize the Horowitz-Maldacena proposal \cite{Horowitz:2003he}? In order to realize this, we need to include matter degrees of freedom (such as Hawking radiation modes) coupled to $\hat{\Psi}(X,Y)$ (for example, see \cite{Brahma:2021xjy}). In addition, the projection must be related to maximally entangled pairs between the star interior and the incoming Hawking radiation, which requires further justification. This goes beyond the scope of the paper, but it is worth asking whether the Horowitz-Maldacena projection will be naturally implemented in this third-quantization framework (although there might be some difficulties, e.g., \cite{Lee:2020aft}).

    \item[--] \textit{Contribution of Euclidean wormholes}: Building on the previous question, we need to investigate a dynamical mechanism that connects the Euclidean void to the coherent state. This dynamical mechanism might be realized by non-linear source terms in the third-quantized action, and its interpretation may relate to interactions between WDW wave functions, which can be well approximated by Euclidean wormholes~\cite{Coleman:1988tj}. If baby-universe $\alpha$-eigenstates dynamically generate the effective source $J(X,Y)$, the statistical distribution of $\alpha$-parameters may determine the coherent-state profile and potentially induce quantum fluctuations or superpositions of the effective horizon scale. This may imply that Euclidean wormholes can contribute to the source term of the third-quantized action; further, this source term induced by Euclidean wormholes may be connected to the active information-transfer mechanism to the outside \cite{Chen}, as the Horowitz-- Maldacena projection works as a quantum teleportation mechanism.

    \item[--] \textit{Connecting to infall problem}: Thanks to third quantization, wormhole interactions can exist between different WDW wave functions, so a given WDW wave function with a classical trajectory should be superposed with alternative trajectories with the same event horizon. This effect may cause horizon-scale uncertainty, partly induced by Hawking radiation and partly by Euclidean wormholes \cite{Lin:2025jmz}. This may reveal the full quantum-gravitational nature inside the horizon, and this issue may relate to the infall problem in the information-loss paradox.
    
    \item[--] \textit{Extensions beyond Schwarzschild}: Applying the same superspace Wick rotation and third-quantized vacuum condition to charged (Reissner--Nordstr\"om) \cite{Chien:2025tzm}, rotating (Kerr), or loop-quantum-gravity-modified black hole interiors \cite{Bouhmadi-Lopez:2020oia} will test whether the Euclidean void boundary condition generically eliminates Cauchy-horizon instabilities as well as spacelike singularities.
\end{itemize}
We leave these questions for future research projects.

\newpage

\section*{Acknowledgment}
The author was supported by the National Research Foundation of Korea (NRF) grant funded by the Korean government (No.~RS-2026-25476711).

\end{document}